\documentclass[conference]{IEEEtran}
\IEEEoverridecommandlockouts
\usepackage{cite}
\usepackage{amsmath,amssymb,amsfonts}
\usepackage{algorithmic}
\usepackage{graphicx}
\usepackage{textcomp}
\usepackage{xcolor}

\usepackage{tikz,xcolor,hyperref}
\usepackage{makecell}
\usepackage[absolute,showboxes]{textpos}
\usepackage{framed}

\definecolor{lime}{HTML}{A6CE39}
\DeclareRobustCommand{\orcidicon}{
	\begin{tikzpicture}
	\draw[lime, fill=lime] (0,0) 
	circle [radius=0.16] 
	node[white] {{\fontfamily{qag}\selectfont \tiny ID}};
	\draw[white, fill=white] (-0.0625,0.095) 
	circle [radius=0.007];
	\end{tikzpicture}
	\hspace{-2mm}
}

\foreach \x in {A, ..., Z}{%
	\expandafter\xdef\csname orcid\x\endcsname{\noexpand\href{https://orcid.org/\csname orcidauthor\x\endcsname}{\noexpand\orcidicon}}
}

\TPMargin{5pt}
\newcommand{\copyrightstatement}{
    \begin{textblock}{0.84}(0.08,0.95)    
         \noindent
         \footnotesize
         \copyright 2021 IEEE. Personal use of this material is permitted. Permission from IEEE must be obtained for all other uses, in any current or future media, including reprinting/republishing this material for advertising or promotional purposes, creating new collective works, for resale or redistribution to servers or lists, or reuse of any copyrighted component of this work in other works.
         DOI: \href{https://ieeexplore.ieee.org/document/9720616}{10.1109/IEDM19574.2021.9720616}
    \end{textblock}
}

\def\BibTeX{{\rm B\kern-.05em{\sc i\kern-.025em b}\kern-.08em
    T\kern-.1667em\lower.7ex\hbox{E}\kern-.125emX}}

\begin{document}

\copyrightstatement

\title{Restructuring TCAD System:\\
Teaching Traditional TCAD New Tricks
}

\author{\IEEEauthorblockN{Sanghoon Myung\orcidA{}\textsuperscript{1,*} , Wonik Jang\textsuperscript{1}, Seonghoon Jin\textsuperscript{2}, Jae Myung Choe\textsuperscript{1}, Changwook Jeong\textsuperscript{1, $\dagger$}, and Dae Sin Kim\textsuperscript{1}}
\IEEEauthorblockA{\textsuperscript{1}Data and Information Technology Center, Samsung Electronics. \textsuperscript{2}Device Lab, Samsung Semiconductor Inc. \\
email: \{shoon.myung\textsuperscript{*}, chris.jeong\textsuperscript{$\dagger$}\}@samsung.com}
}

\maketitle
\begin{abstract}
Traditional TCAD simulation has succeeded in predicting and optimizing the device performance; however, it still faces a massive challenge — a high computational cost. There have been many attempts to replace TCAD with deep learning, but it has not yet been completely replaced. This paper presents a novel algorithm restructuring the traditional TCAD system. The proposed algorithmpredicts three-dimensional (3-D) TCAD simulation in real-time while capturing a variance, enables deep learning and TCAD to complement each other, and fully resolves convergence errors.
\end{abstract}


\section{Introduction}
Technology computer-aided design (TCAD) simulation has played a key role in predicting and optimizing semiconductor device performance. As semiconductor technologies have become more sophisticated, TCAD simulation gets more time-consuming. To reduce simulation time, there have been attempts to use some technologies \cite{b1,b2} based on multi-core computing, which turned out to be insufficient. As an alternative way, researchers began introducing deep learning (DL) models. In the semiconductor field, two approaches have been mainly studied. The first one is to model the relationship between inputs and outputs using DL rather than using partial differential equations (PDE) \cite{b3,b4,b5,b6}. The second one is to solve differential equation via DL, which provides the initial values for PDE \cite{b7} or regularizes itself to mimic the differential equation \cite{b8}. So far no study has been done to make DL and current TCAD model compatible with each other. In this paper, we present a novel algorithm, called Real-Time TCAD (RTT) to complement the TCAD simulation process. This proposed algorithm enables 3-D real-time TCAD simulation and allows both TCAD and DL to be compatible with each other. As a result, the new features of this algorithm would restructure the conventional TCAD system in a way that DL solves the troubles that TCAD has and vice versa.

\section{METHODOLOGY AND ITS ASSESSMENT}

\subsection{Preliminary}
\begin{figure*}[t]
    \centering
        \begin{minipage}{.38\linewidth}
            \includegraphics[width=\linewidth]{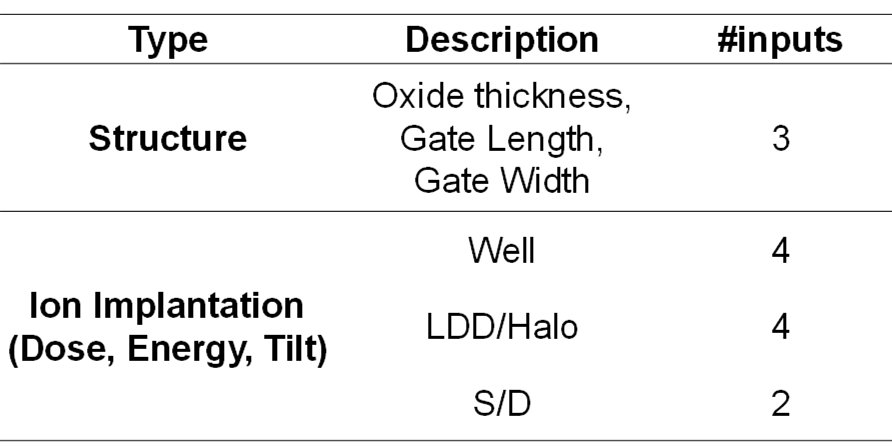}
            \caption{Description of TCAD input parameters.There are three variables related to structure and ten variables associated with ion implantation.}
            \label{Fig1}
        \end{minipage}
        \begin{minipage}{.6\linewidth}
            \includegraphics[width=\linewidth]{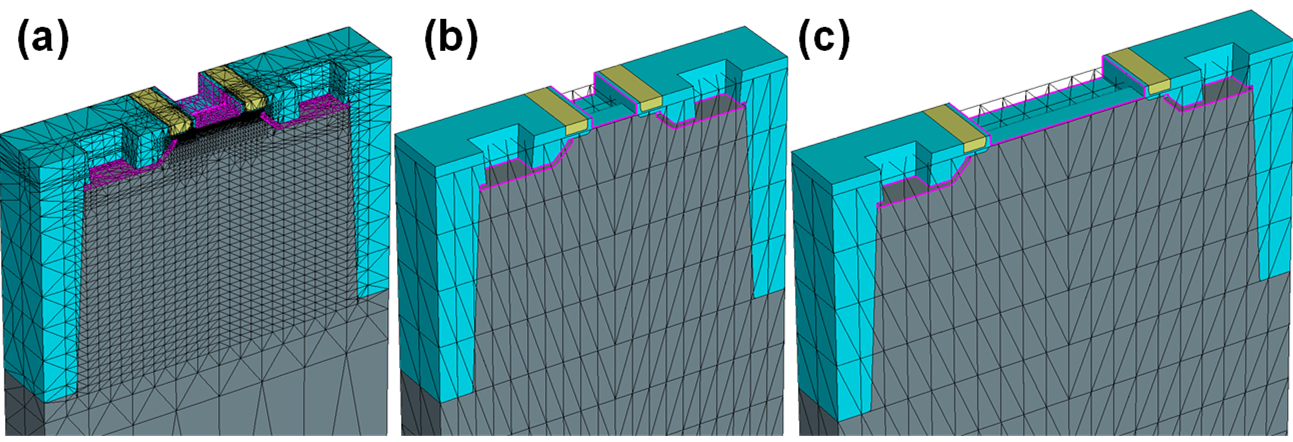}
            \caption{An example of (a) original mesh, (b) interpolated mesh that has a same structure of (a) and (c) interpolated mesh that has a long gate length. To make it clearly visible, we enlarge interpolated mesh bigger than the actual scale. Regardless of structure size, the number of grids is constant while the size of the interval depends on the structure size.}
            \label{Fig2}
        \end{minipage}
\end{figure*}

We begin with a brief explanation about TCAD simulation on a 45nm process. We use in-house process and device simulator (Polaris). There are three and ten input variables on device geometry and ion implantation (IIP) process, respectively (Fig. \ref{Fig1}). We change the unstructured mesh into structured meshes with uniform intervals since it is difficult for DL to learn an unstructured mesh (Fig. \ref{Fig2}). Even if the size of device varies depending on the input conditions, we change them into the predefined size (Fig. \ref{Fig2}). Then, we generate 2,050 TCAD simulation samples by changing the input variables. 2,000 samples are used in the training phase and the remaining 50 samples are used to evaluate the performance of models.
\subsection{RTT Process Model}
\begin{figure*}[h!]
    \centering
    \begin{minipage}{.47\linewidth}
        \includegraphics[width=\linewidth]{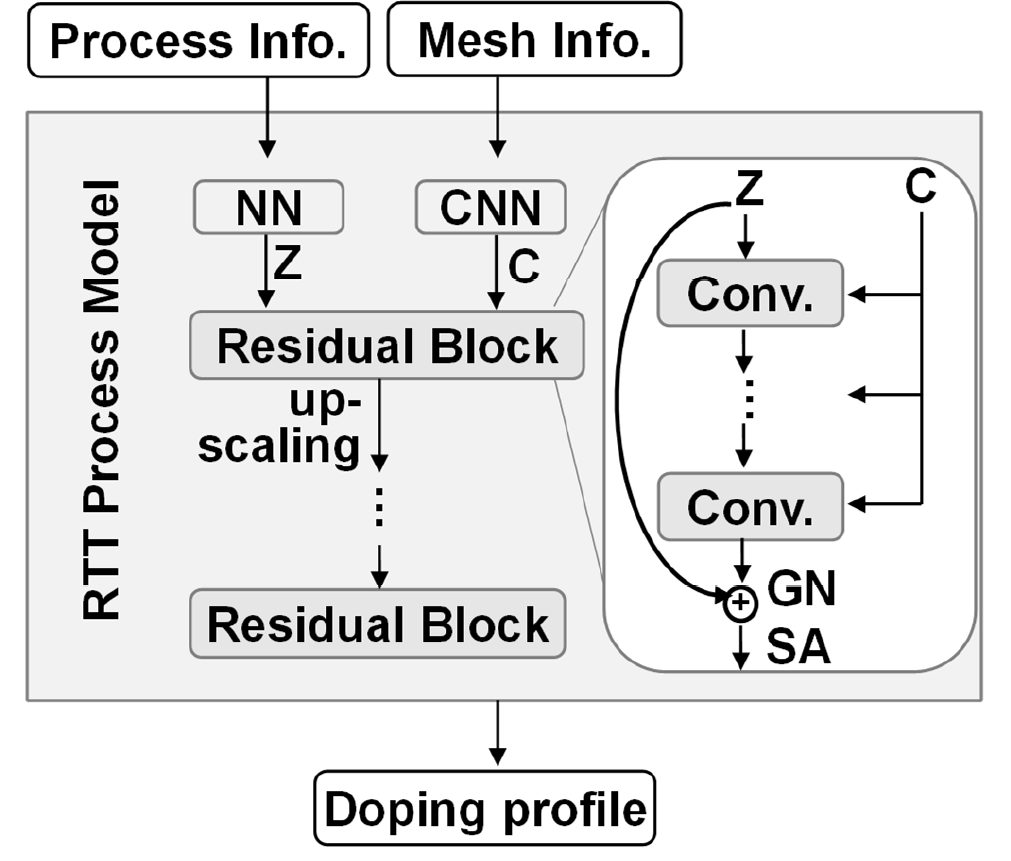}
        \caption{Illustration of proposed process model. NN denotes a neural network, Z the latent space learned by NN, C the coordinates sampled from mesh, GN the group normalization and SA the swish activation. Z and C are concatenated for every convolution.Up-sampling doubles the dimensions of input representation.}
        \label{Fig3}
    \end{minipage}
    \begin{minipage}{.48\linewidth}
        \includegraphics[width=\linewidth]{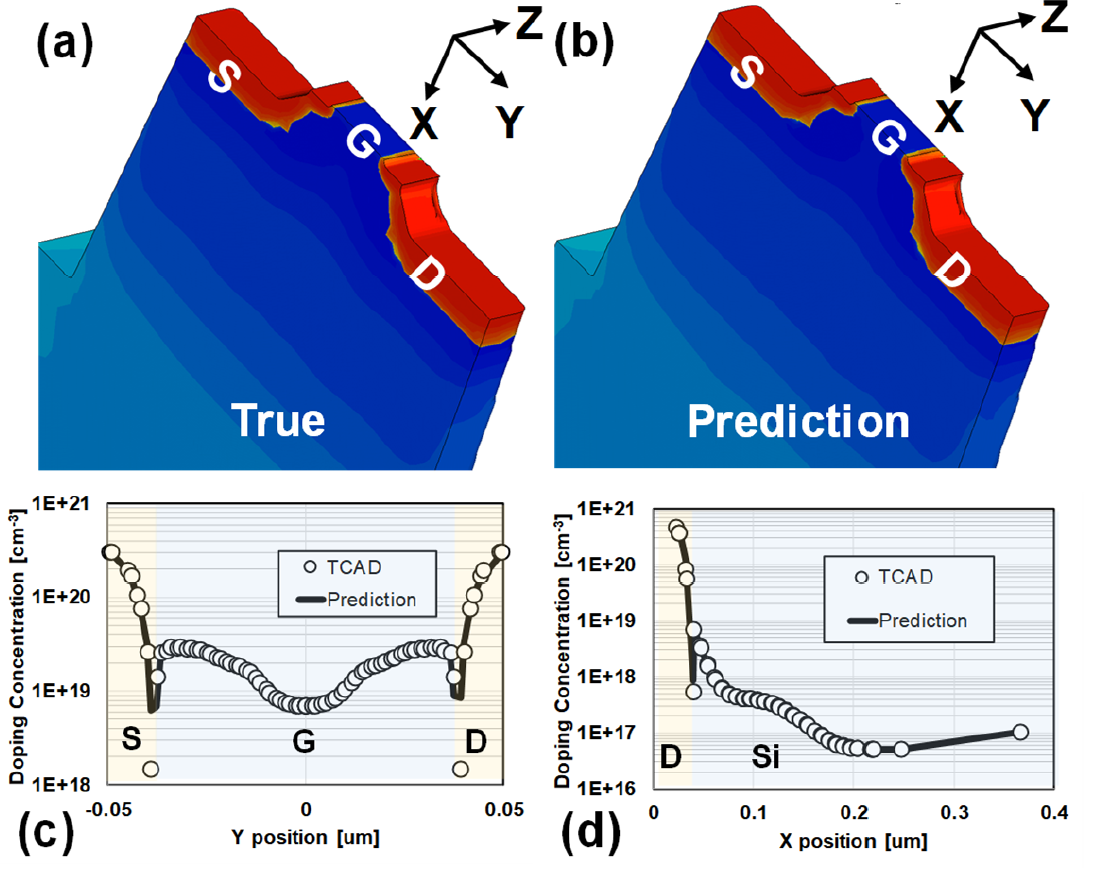}
        \caption{(a) A TCAD process simulation result. (b) A prediction result of RTT process model. (c) 1-D doping concentration plot in the horizontal direction below the gate. (d) 1-D doping concentration plot in the vertical direction at the center of drain.}
        \label{Fig4}
    \end{minipage}
\end{figure*}

Process simulation computes doping concentration in metal-oxide-semiconductor field-effect transistor (MOSFET) with respect to process condition. The process simulation consists of two steps; the first step creates MOSFET structure with meshes, and the second step solves IIP and diffusion processes using PDE. We use DL to primarily focus on predicting the doping profile since the latter step takes most of the time during process simulation.The distinctive feature of our work compared to the previous ones is that the process model learns the real value of doping of MOSFET structures, not the images. Furthermore, it allows the model to handle 3-D doping profiles. We introduce a core architecture based on convolutional neural network (CNN), up-sampling and residual block (RB) (Fig. \ref{Fig3}). Up-sampling doubles the dimensions of input representation. RB, which contains group normalization (GN) \cite{b9} and swish activation (SA) \cite{b10}, is introduced to prevent vanishing gradients \cite{b11}. GN makes the training loss landscape smoother and SA avoids a slow training time during near-zero gradients. 
Last but not least, putting coordinates during performing convolution operations is very crucial for model performance. The previous studies \cite{b3,b12} have put relative coordinates in each CNN operation. On the other hand, in this study, the coordinates extracted by actual meshes must be inserted because the coordinates of uniformed-sized outputs could be different. (Fig. \ref{Fig2} (b), (c)). We concatenate the Cartesian coordinates channel-wise to the input representation after passing the adaptive network, which consists of CNN and pooling layer. An adaptive network is designed to match the size with which each RB deals. Figure \ref{Fig4} compares the prediction results of the RTT process model (Fig. \ref{Fig4} (b)) to the ground truth (Fig. \ref{Fig4} (a)). To compare TCAD with the RTT model more precisely, we assess 1-D doping profiles exploited by horizontal and vertical axes (Fig. \ref{Fig4} (c), (d)). RTT model has achieved the average accuracy above 99\% compared to TCAD.

\subsection{RTT Device Model}
\begin{figure*}[h!]
    \centering
    \begin{minipage}{.26\linewidth}
        \includegraphics[width=\linewidth]{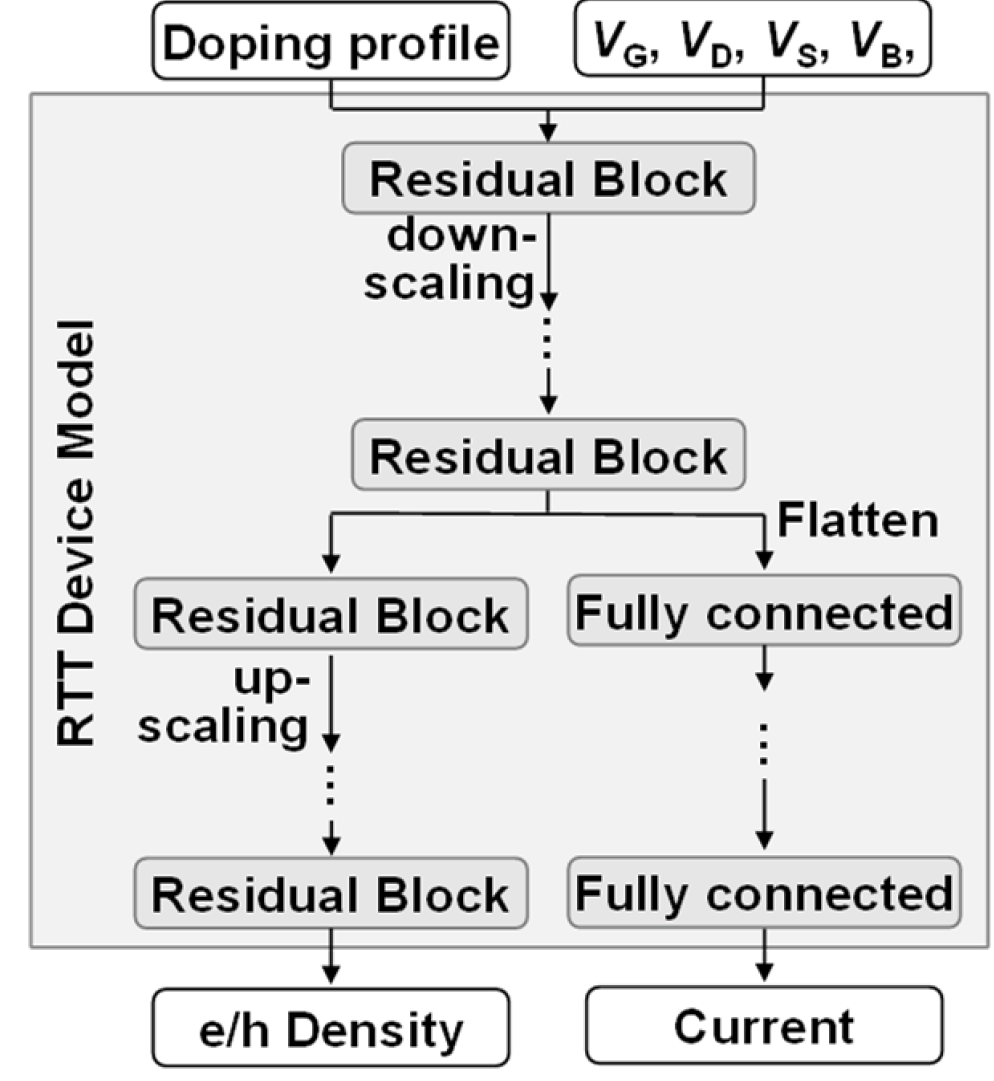}
        \caption{Illustration of RTT device model, which can predict both electron/hole density and the current.}
        \label{Fig5}
    \end{minipage}
    \begin{minipage}{.65\linewidth}
        \includegraphics[width=\linewidth]{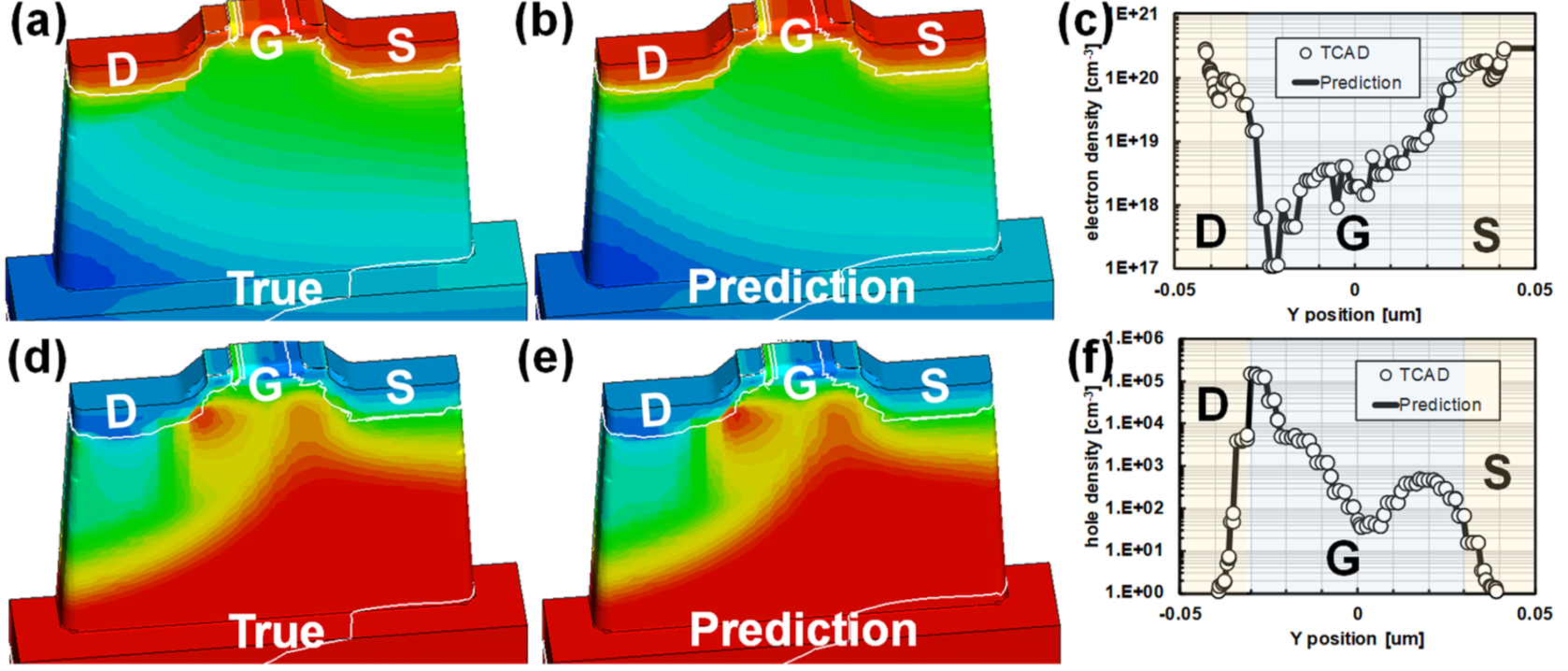}
        \caption{Electron density of (a) TCAD (b) RTT device model at $V_D=V_G=V_{DD}$. (c) 1-D electron density in the horizontal direction below the gate. Hole density of (d) TCAD (e) RTT device model at $V_D=V_G=V_{DD}$. (f) 1-D hole density in the horizontal direction below the gate.}
        \label{Fig6}
    \end{minipage}
\end{figure*}

In device simulation, MOSFET structure and its doping concentration from RTT process model work as input variables. The device simulation results in the current-versus-voltage ($I-V$) curve and carrier profiles. We propose a multi-task learning that predicts the current and carrier profile simultaneously. Inputs to RTT device model, MOSFET structure with doping concentration and bias conditions, share RB and down-sampling until they reach the diverging point (Fig. \ref{Fig5}). The network for career density uses RB and up-sampling for outputs to become bigger, same-sized with input dimension; the network for current only uses fully connected layers. Both electron and hole profiles predicted by RTT device model are identical with TCAD simulation by 99.5\% (Fig. \ref{Fig6}).

\section{Additional Features}
\vspace{10pt}
\subsection{Real-time Simulation and No Convergence Error}
\begin{figure*}[t!]
    \centering
    \begin{minipage}{.32\linewidth}
        \includegraphics[width=\linewidth]{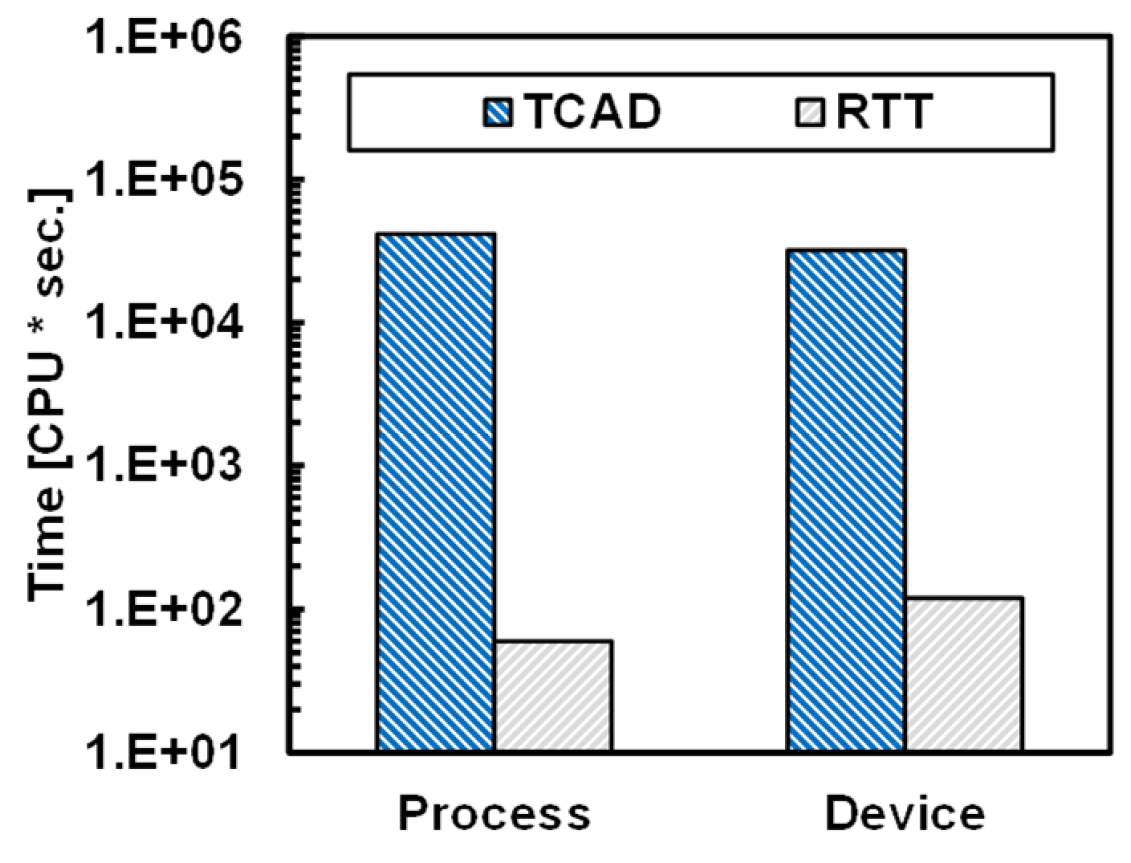}
        \caption{Prediction time for each simulation. RTT model can predict the result up to 691 times faster.}
        \label{Fig7}
    \end{minipage}
    \begin{minipage}{.31\linewidth}
        \includegraphics[width=\linewidth]{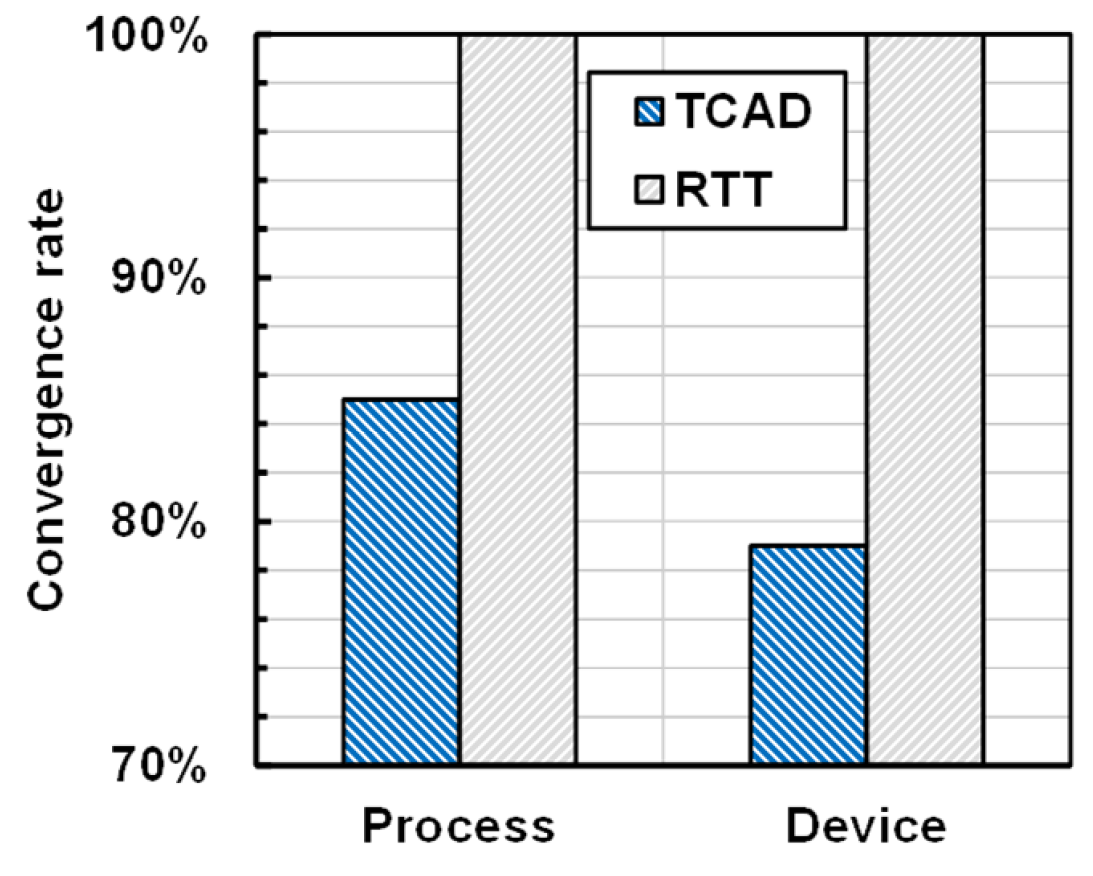}
        \caption{Convergence rate of each simulation. RTT model shows 100\% convergence rate.}
        \label{Fig8}
    \end{minipage}
    \begin{minipage}{.35\linewidth}
        \includegraphics[width=\linewidth]{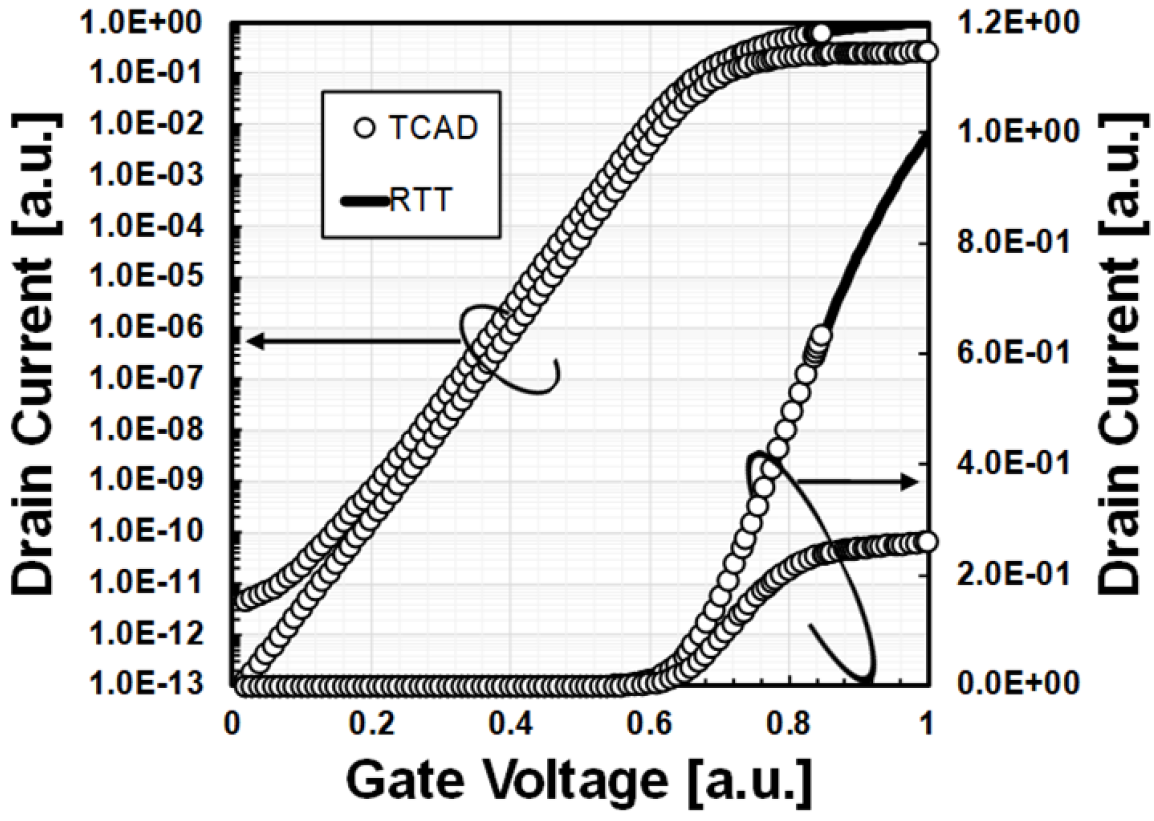}
        \caption{An example of $I_D-V_G$ curves. RTT device model successfully calculates $I_D-V_G$ curves while TCAD simulation cannot.}
        \label{Fig9}
    \end{minipage}
\end{figure*}

One of the merits of RTT models, including RTT process and device model, is that it can predict TCAD simulation in real time. As described in Sec. I, TCAD simulation has trouble in predicting the device performance in real time. In contrast, RTT models can predict it in almost real time (Fig. \ref{Fig7}). Quantitatively speaking, RTT models are 691 times faster than TCAD simulation. On top of that, TCAD simulation often suffers from errors that fail to solve PDE. The longer the TCAD tool chain, the lower the convergence rate (Fig. \ref{Fig8}). On the other hand, RTT models present 100\% convergence results. Figure \ref{Fig9} shows that the RTT device model is successful at predicting the $I-V$ curve, which TCAD fails to predict.
\vspace{10pt}
\subsection{Compatibility with TCAD simulation}
\begin{figure*}[t!]
    \centering
    \begin{minipage}{.36\linewidth}
        \includegraphics[width=\linewidth]{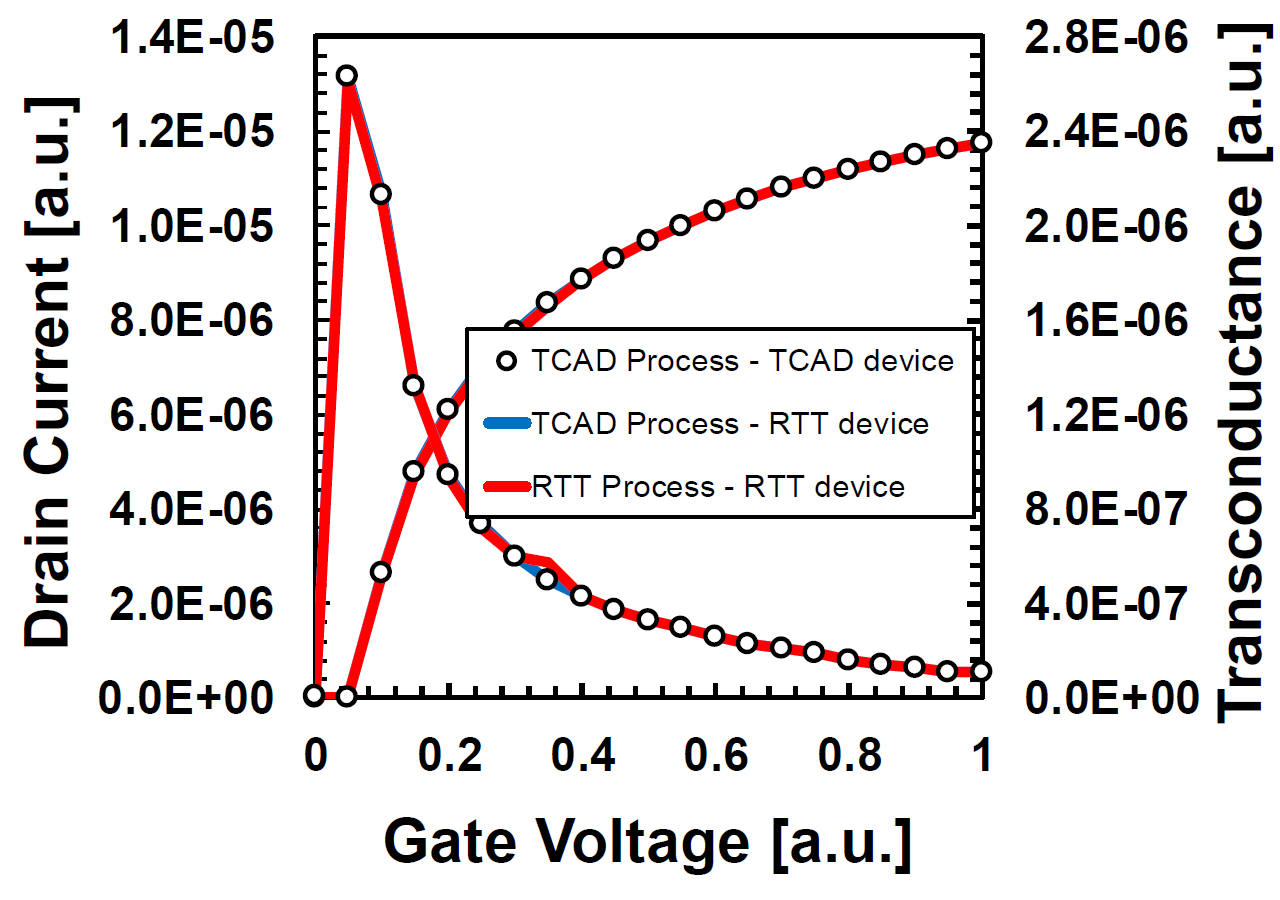}
        \caption{$I_D-V_G$ curves calculated by TCAD device simulation whose inputs are from both TCAD process simulation and RTT process model.}
        \label{Fig10}
    \end{minipage}
    \begin{minipage}{.3\linewidth}
        \includegraphics[width=\linewidth]{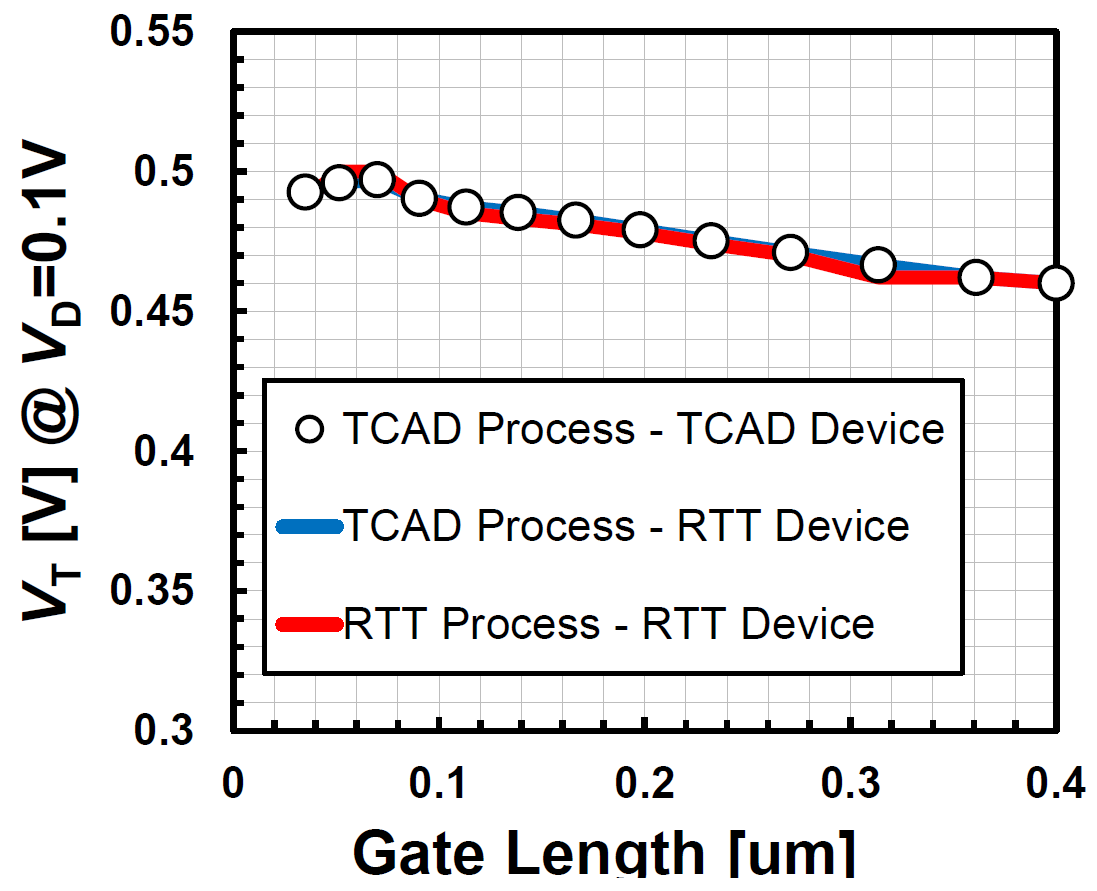}
        \caption{The results of the threshold voltage roll-off by various simulations. The results can be indistinguishable from each other.}
        \label{Fig11}
    \end{minipage}
    \begin{minipage}{.31\linewidth}
        \includegraphics[width=\linewidth]{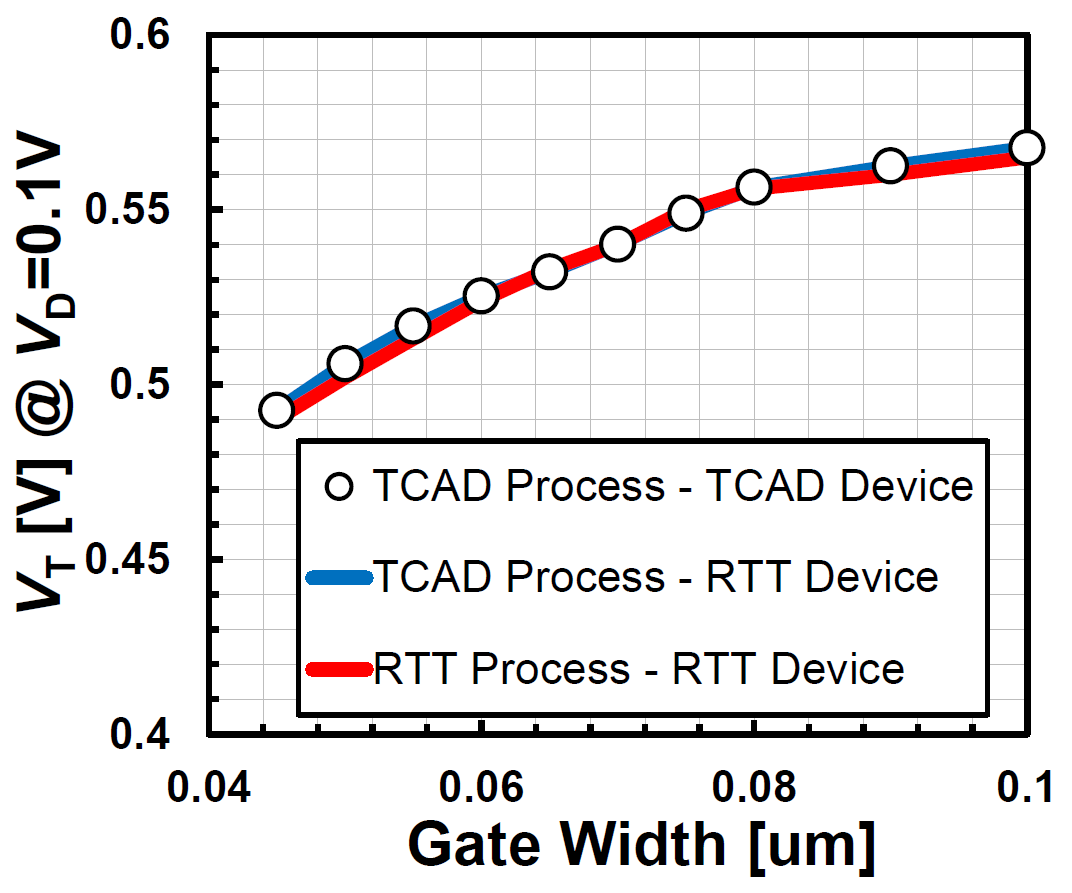}
        \caption{The results of narrow width effect. All results are consistent with each other.}
        \label{Fig12}
    \end{minipage}
\end{figure*}

\begin{figure*}[t!]
    \centering
    \begin{minipage}{.36\linewidth}
        \includegraphics[width=\linewidth]{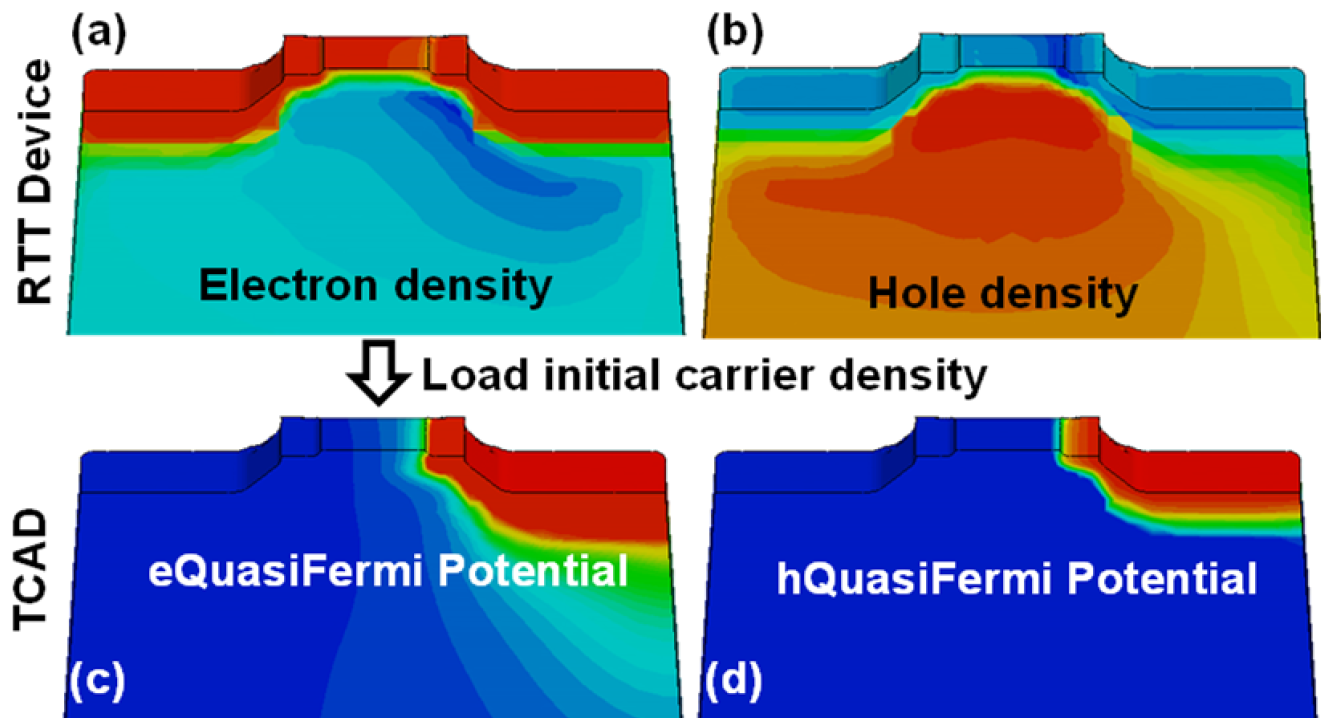}
        \caption{(a) Electron and (b) hole density predicted by RTT. The quasi-Fermi potential of (c) electron and (d) hole estimated by TCAD after loading carrier density from RTT device results.}
        \label{Fig13}
    \end{minipage}
    \begin{minipage}{.3\linewidth}
        \includegraphics[width=\linewidth]{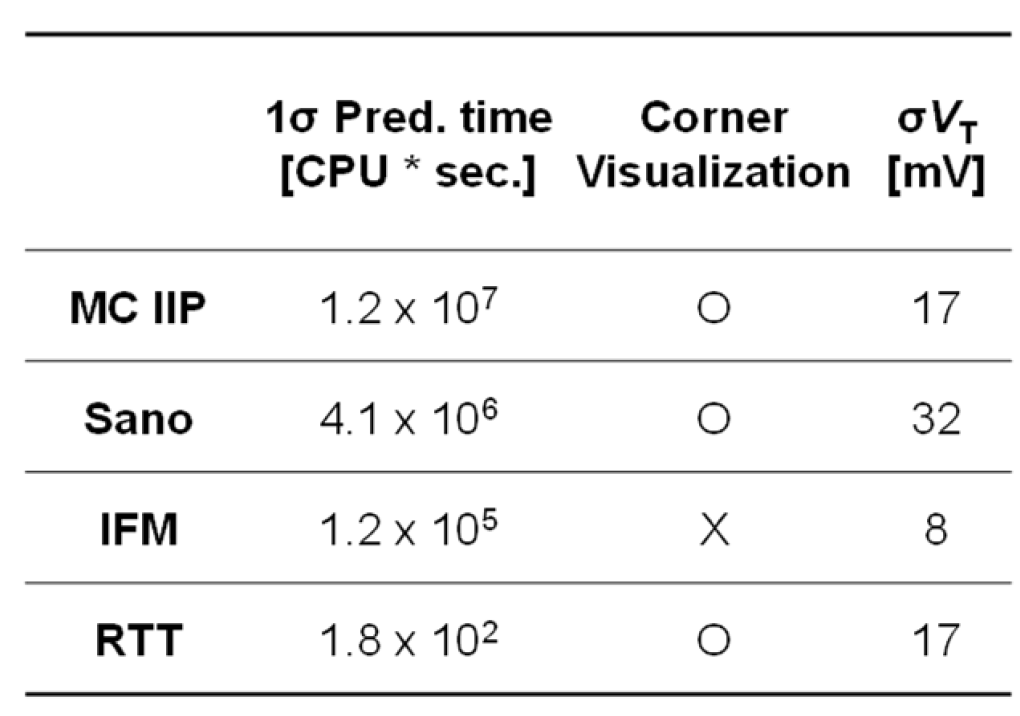}
        \caption{The results of benchmark methods. Prediction time represents the total time to generate a hundred samples. RTT model shows a good agreement with MC IIP results.}
        \label{Fig14}
    \end{minipage}
    \begin{minipage}{.31\linewidth}
        \includegraphics[width=\linewidth]{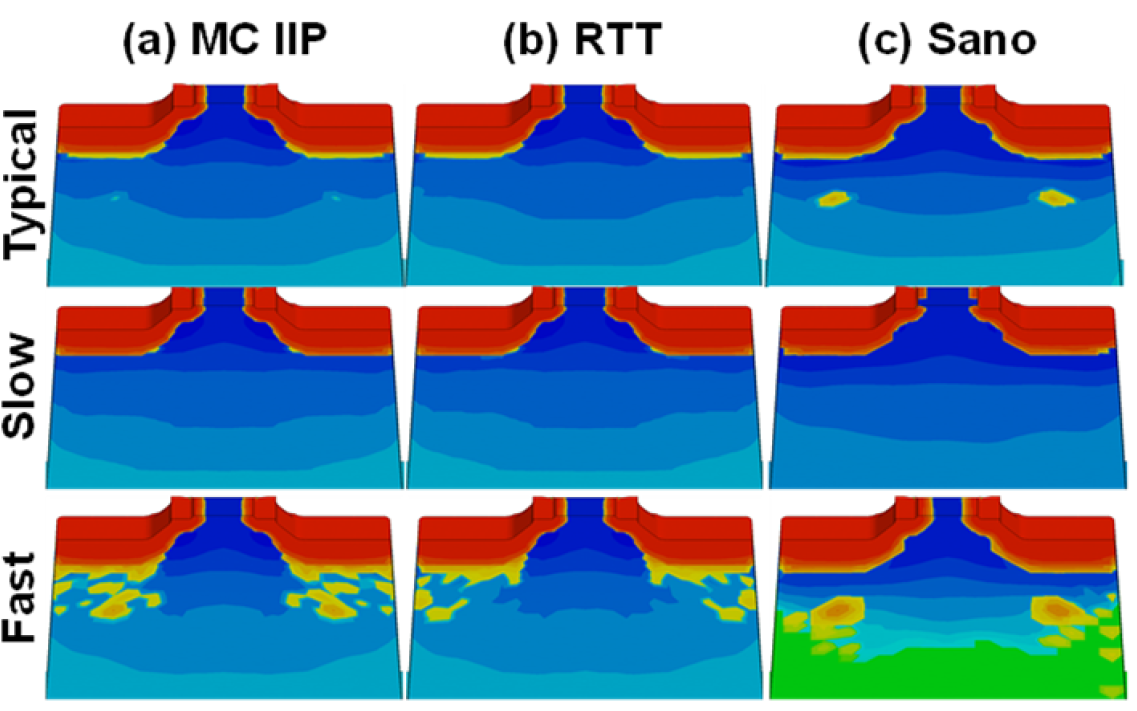}
        \caption{Process corner of (a) MC IIP (b) RTT (c) Sano. Typical $V_T$ denotes an average of 100 samples, Slow $V_T(u_{donor}$ - $3\sigma_{donor}$) - ($u_{acceptor}$ + $3\sigma_{acceptor}$) and Fast $V_T(u_{donor}$ + $3\sigma_{donor}$) - ($u_{acceptor}$ - $3\sigma_{acceptor}$).}
        \label{Fig15}
    \end{minipage}
\end{figure*}

The previous studies \cite{b3,b4,b5} have only predicted the results. On the contrary, the proposed algorithm makes inputs and outputs of TCAD and RTT interchangeable. Figure \ref{Fig10} demonstrates that all results are indistinguishable even if an input of TCAD device simulation is an output of whether RTT process model or TCAD process simulation. In addition, the RTT device model infers $I-V$ characteristics without any difference between the outputs of TCAD and those of RTT. Figure \ref{Fig11} and \ref{Fig12} demonstrate that RTT models are compatible with conventional TCAD model in various situations. Figure \ref{Fig13} shows an example that RTT device model can replace a part of TCAD device simulation. The TCAD device simulator self-consistently couples the Poisson and continuity equation to obtain the unknown variables: carrier density and electrostatic potential. With our RTT device model, however, TCAD device simulator solves only Poisson equation after loading the carrier density predicted by RTT device model (Fig. \ref{Fig13}). This method enables TCAD device simulator to use a Schottky contact model \cite{b13} while the previous work \cite{b7} cannot.
\vspace{10pt}
\subsection{Statistical Inference}
\begin{figure*}[h!]
    \centering
    \begin{minipage}{.47\linewidth}
        \includegraphics[width=\linewidth]{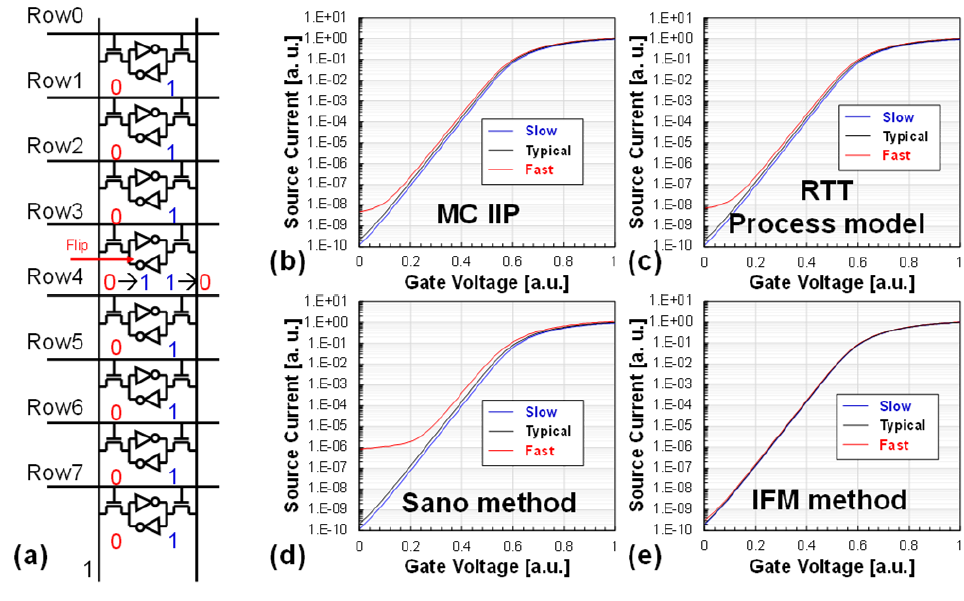}
        \caption{(a) A schematic on failure analysis of SRAM array. At row 3, leakage current occurs from the drain to the source. (b-e) $I_S-V_G$ curve of (b) MC IIP (c) RTT (d) Sano and (e) IFM for the process corner.}
        \label{Fig16}
    \end{minipage}
    \begin{minipage}{.47\linewidth}
        \includegraphics[width=\linewidth]{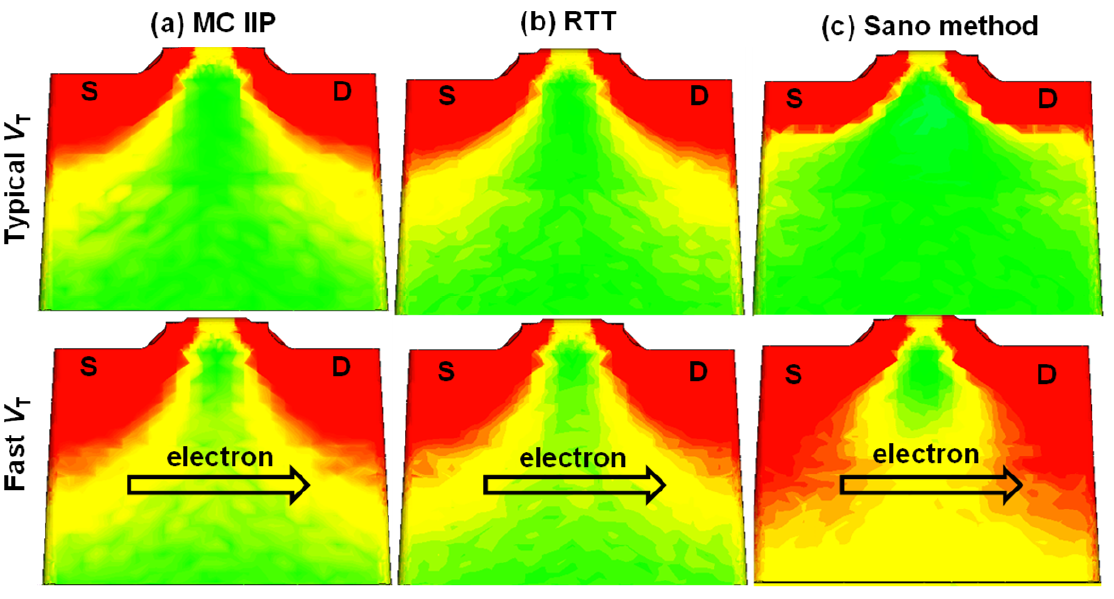}
        \caption{The results of electron current density. $I_{DS}$ is well suppressed in the typical process,whereas unexpected current flows at the substrate in the fast process. All method shows the possibility of the leakage current.}
        \label{Fig17}
    \end{minipage}
\end{figure*}

The doping profile can naturally contain a variation because IIP is usually computed by Monte Carlo (MC) method. To capture the variance of MC IIP, we assume that the doping concentration of each node follows Gaussian distribution. Although discrete dopants are distributed in a device, this assumption is valid because the diffusion model is continuum model. We set the loss function of the RTT process model as negative-log likelihood (NLL) to capture a variance.
\begin{equation}
NLL=\frac{(y-u(x))^2}{2\sigma^2(x)}+\frac{\ln{\sigma^2(x)}}{2}\label{eq}
\end{equation}
where $u(x)$ and $\sigma^2(x)$ are mean and variance of doping concentration that RTT process model returns. The $\sigma^2(x)$ can represent the data noise (e.g., variance of MC IIP) if there is enough data \cite{b14}. Note that the number of MC particles for IIP is calculated from a dose of IIP and structure dimension.
We evaluate whether the proposed method captures a variation of MC IIP or not. Furthermore, we compare MC IIP with two existing methods: Sano method \cite{b15} and impedance Field Method (IFM) \cite{b16}. The RTT results are compared to the various benchmark methods (Fig. \ref{Fig14}). The $\sigma V_T$ of RTT is consistent with that of MC IIP and is approximately an average of the $\sigma V_T$ obtained from Sano and IFM. Also note that RTT model provides good visualization for simulation results. Figure \ref{Fig15} shows the process corner results from each method, but IFM cannot illustrate the process corner. RTT is consistent with MC IIP results for process corner. RTT and MC IIP results are similar to Sano results except for the fast corner. Next, to check the impact of the process variability on SRAM failure, simulations at various process corners have been performed and analyzed (Fig. \ref{Fig16} and Fig. \ref{Fig17}). Figure \ref{Fig16} (a) illustrates a test case where a defective cell in one of SRAM arrays causes a punch-through current. Process corner simulations with different methods (RTT, Sano) are compared to MC IIP in order to examine which approach correctly describes a variation of doping profile. Except for the IFM method, the other methods (MC, Sano, and RTT) show the possibility of the leakage current (Fig. \ref{Fig16}). It can be clearly seen in Fig. \ref{Fig17} that unexpected current flows on the substrate for the fast corner condition, while the typical condition suppresses the leakage current. It is important to note that RTT model is almost $10^3\sim10^5$ times faster than MC and Sano when predicting 1$\sigma$ doping profile.
\vspace{10pt}
\section{Conclusion}
We propose a novel approach to reinforce traditional TCAD simulation with domain-tailored DL algorithms. The proposed RTT method 1) shows the real-time prediction of 3-D TCAD simulators for the first time, 2) enables TCAD and deep learning to be compatible with each other, 3) fully resolves convergence error issues, and 4) demonstrates accurate modeling of IIP-induced process variations with $\>10^3$x reduction in simulation time. We hope this work can facilitate future studies of restructuring conventional TCAD systems.
\newpage

\end{document}